\documentclass[twoside]{article}
\usepackage{graphicx}
\usepackage{fancyhdr}
\usepackage{multicol}
\usepackage{styl-ap}

\begin{document}
\title{X-ray observations of VY Scl type nova-like binaries in the high and low state}
\author{Polina Zemko\work{1,2}, Marina Orio\work{2,3}}
\workplace{Sternberg Astronomical Institute, Moscow State University, 13, Universitetskij ave., Moscow, 119991, Russia
\next
INAF - Osservatorio di Padova, vicolo dell' Osservatorio 5, I-35122 Padova, Italy
\next
Department of Astronomy, University of Wisconsin, 475 N. Charter Str., Madison, WI 53704, USA}
\mainauthor{polina.zemko@gmail.com}
\maketitle

\begin{abstract}
Four VY Scl-type nova-like systems were observed in X-rays both during the low and the 
high optical states. They are BZ Cam, MV Lyr, TT Ari, and V794 Aql.
Using archival {\sl ROSAT}, {\sl Swift} and {\sl SUZAKU} observations we found that the 
X-ray flux for BZ Cam is higher during the low state, but there is no supersoft X-ray 
source (SSS) that would indicate the thermonuclear burning predicted in a previous article.
 The X-ray flux is lower by a factor 2-10 in the low than the high state in other 
 systems, and does not reflect the drop in $\dot{M}$ inferred from optical and UV data. 
 The best fit model for the X-ray spectra is a collisionally
 ionized plasma model. The X-ray flux may originate in a shocked
 wind or in accretion onto polar caps in intermediate polar systems that continues
 even during the low state.
 
\end{abstract}


\begin{multicols}{2}
\section{Introduction}

Nova-like (NLs) stars are non-eruptive CVs (Warner, 1995), classified into several 
subtypes according to their properties. Here we will focus on the VY 
Scl-type NLs or ``anti-dwarf novae'' characterized by the presence of 
occasional dips on the light curve, so-called low states.

The large optical and UV luminosity has suggested that in the high state these objects are
 undergoing mass transfer onto the WD at a high rate $\dot{M}$, 
 $>10^{-10}$M$_{\odot}$ year$^{-1}$, sustaining an accretion disk
 in a stable hot state and preventing dwarf novae (DNe) outbursts. 
 The low states have been attributed to a sudden drop of the $\dot{M}$
 from the secondary, or even to a total cessation of a mass transfer (Hessman, 2000). 

Greiner et al. (1999) proposed a link between the VY Scl-type stars and super
soft X-ray sources based on a {\sl ROSAT} observations of V751 Cyg. These authors 
found an anti-correlation in the optical and X-ray intensity, and despite
 the very poor spectral resolution of the {\sl ROSAT} HRI, 
 the spectrum appeared to be very soft in the low state. 
These authors suggested that quasi-stable thermonuclear burning occurs on the surface of 
the WD in the low state. In this framework, 
VY Scl-type stars are key objects in the evolution of interacting WD binaries, in which
 hydrogen burning occurs periodically without outbursts. Thus they may reach the
 Chandrasekhar mass and the conditions for type Ia supernovae explosion.

Using archival X-ray observations, in this paper we compare the high and low state X-ray 
data of 4 VY Scl-type stars attempting to reveal evidence of nuclear burning during the 
low state, or seeking alternative explanations for the changes that take place during the 
 transition from the high to low state.

\section{Observation and Data analysis}

We examined the archival X-ray data of VY Scl-type stars obtained with {\sl Swift}, 
{\sl ROSAT} and {\sl SUZAKU} and chose the objects that were observed both 
in the high and low states: BZ Cam, MV Lyr, TT Ari and V794 Aql. 
 To determine when the low and high optical states occurred, we relied on the data of the 
 Variable Star Network (VSNET) Collaboration (Kato et al., 2004), AAVSO \footnote{ 
http://www.aavso.org} and ASAS databases. We used HEASOFT Version 6-13 to analyze the data
 and XSPEC version 12.8.0 for spectral modeling. We also estimated the UV magnitudes
 of the objects using both {\sl Swift}/UVOT and additional {\sl GALEX} archival observations.

\section{Results}
In Figure 1 we show
the high and low state X-ray spectra of 
BZ Cam, MV Lyr, TT Ari and V794 Aql and the fits
 to these spectra. The fit parameters are
summarized in Table 1 and 2. 

\begin{myfigure}
\begin{center}
\centerline{\resizebox{90mm}{!}{\includegraphics{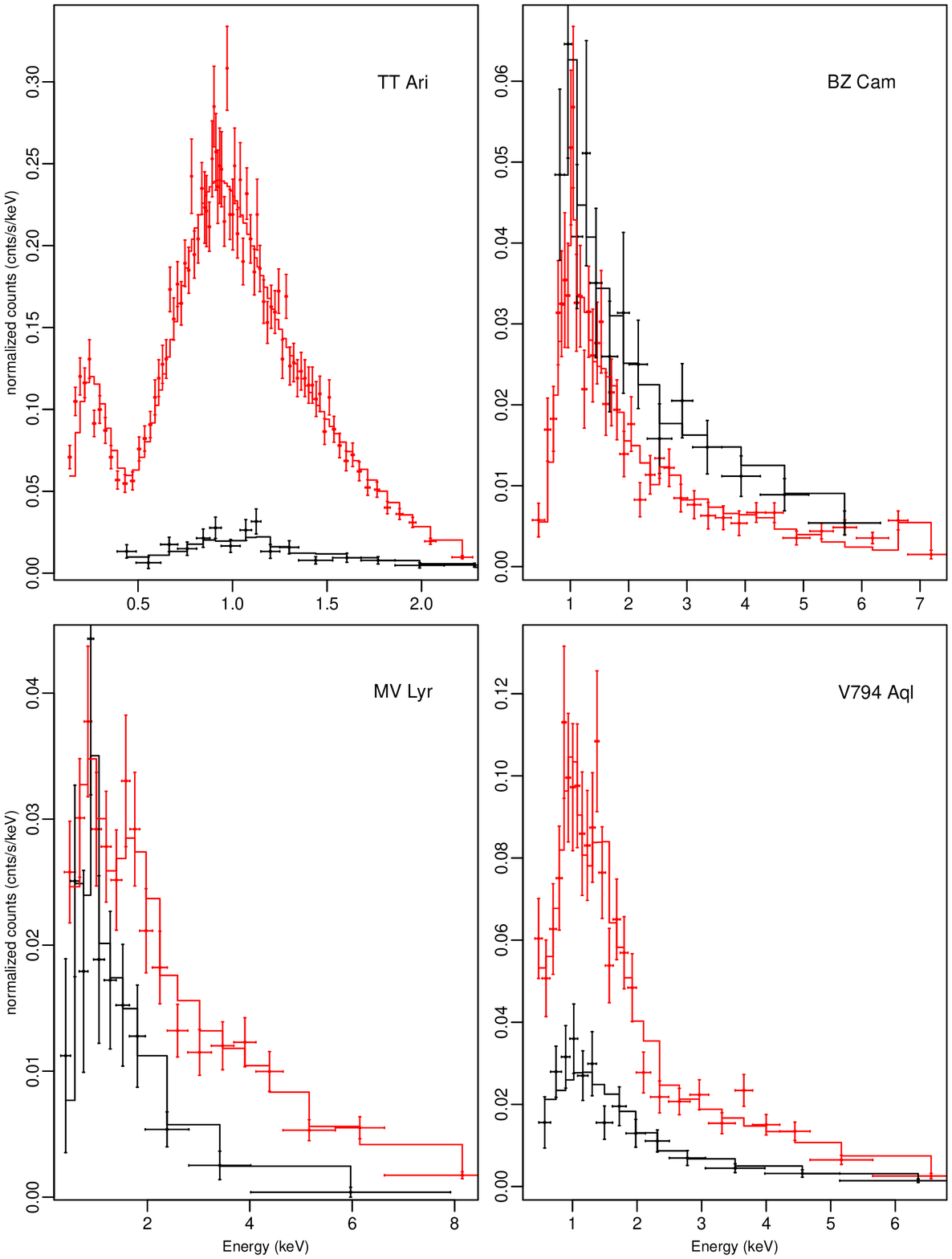}}}
\caption{\footnotesize{The low and high states X-ray spectra of BZ Cam, MV Lyr and V794 Aql 
observed with {\sl Swift}. TT Ari was observed with {\sl Swift} during the low state and 
with {\sl ROSAT} during the high state. The high state spectra are plotted
 in red and the low state spectra 
in black.
The solid lines show to spectral fits, the dots with error bars
 are the data points.}}
\end{center}
\end{myfigure}

TT Arietis (top left panel) is one of the most optically
luminous CVs,
 usually at mag 10-11. Sometimes it abruptly falls into an ``intermediate state'' around 
 14 mag. or even into a ``low state'' at about 16.5 mag. According to 
Belyakov et al. (2010), this binary system consists of a
0.57-1.2 M$_{\odot}$ white 
dwarf and a 0.18-0.38 M$_{\odot}$ secondary component.
The
orbital period is $P_{\rm{orb}}=0.13755114$ d. (Thorstensen,  Smak \& Hessman, 1985).
The high state X-ray flux appeared to be about twice larger than in the low state. 
We fitted the high state spectrum with a two components thermal plasma model (APEC model
 -- emission spectrum from collisionally ionized diffuse gas calculated using the ATOMDB 
 code v2.0.1), and also with one component black body model.
The low state spectrum is also well fitted with a 2 components thermal plasma model. 

BZ Cam is a NL star at a distance $830 \pm 160~pc$ (Ringwald\&Naylor, 1998) with
an orbital period of 0.153693(7) d. (Patterson et al., 1998). Most of the time
it shows brightness variations around V = 12 - 13 mag., with rare occasional transitions 
to the 14 - 14.5 mag. low states. 
BZ Cam is surrounded by a bright emission nebula with a bow-shock structure.
Hollis (1992) proposed that the bow shock structure must be
 due to the interactions of wind from BZ Cam with the interstellar medium. 
Greiner et al. (2001) suggested that this nebula is photoionized by a bright 
 central object that must be a super soft X-ray source, while the
 bow shock structure is due to the high proper motion of BZ Cam.

From the second plot of Fig. 1 it can be seen that the luminosity 
is higher in the low state, however, in the very soft spectral region,
at energy $\leq$0.5 keV, the X-ray flux is almost twice higher in the high state,
 which is exactly the opposite of the idea proposed
 by Greiner et al (1999). Interestingly, the spectral fits in both states 
indicate that we may be observing a strong Ne X Lyman${\alpha}$ line at 1.02 keV. 

We fit the low state spectrum of BZ Cam, with a two-component APEC model.
The high state of BZ Cam is best fitted 
with a one component Raymond-Smith model of hot, diffuse gas 
or with a two components APEC model. 
In this fit we found an underabundance (with respect to solar values) of 
 C and N and and an overabundance of O, Ne, and intermediate mass elements like S and Ca. 

\begin{myfigure}
\centerline{\resizebox{60mm}{!}{\includegraphics{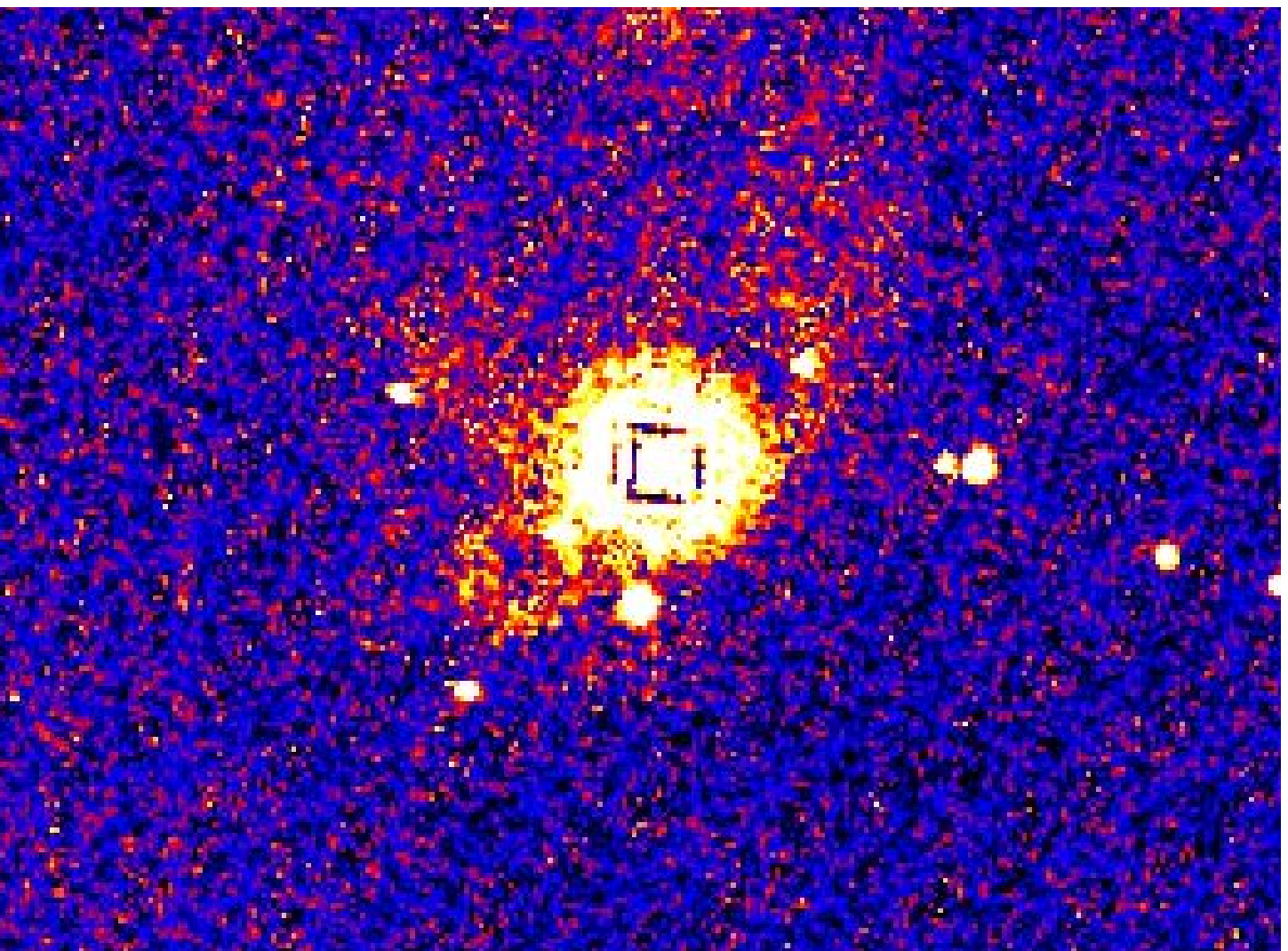}}}
\caption{The {\sl Swift} UV image of BZ Cam, in which the bow-shock emission nebula
 is clearly detected.}
\end{myfigure}

The UV magnitudes of BZ Cam in the high and low states indicate a smaller variation 
than observed in the other objects (albeit in different UV filters).
 This is explained by Figure 2 in which we show the UV image of
 the nebula obtained with {\sl Swift}/UVOT observations. 
Obviously the ionized nebula also emits copious UV flux. Comparing the image of BZ Cam in 
Figure 2 and the one, presented in the Figure 4 in (Greiner et al., 2001), one 
can see an additional prominent feature of the nebula in the former image. In UV there is 
a bright ring in the bottom-left side of the BZ Cam nebula that was absent in O III and
 H$_{\alpha}$ in September 2000.

MV Lyr is 
at about 12th-13th mag in the high state and 16th-18th mag in the low state. 
Hoard et al. (2004) have shown that the distance to the object is
 $505\pm50$ pc. 
With their {\sl FUSE} (Far Ultraviolet Spectroscopic Explorer) observations these authors 
estimated an upper limit to $\dot{M}$ during the 
low state $\leq 3\times10^{-13} M_{\odot}/yr$. The orbital period of this system
is 3.19 h (Skillman, Patterson \& Thorstensen, 1995).

 Our data show that the high state X-ray flux of MV Lyr is 
higher by an order of magnitude than in the low state. The spectrum
 is also harder with additional component prominent above 1.7 keV.
 We fitted the high state spectrum of MV Lyr with a two components thermal plasma model,
 A good fit is also obtained with a thermal plasma and a power law model. The low state 
 spectrum was also fitted with a two components plasma model (see Table 1).  

\begin{table*}
	\footnotesize
	\begin{center}
 		\caption{Fitting models and parameters for BZ Cam and MV Lyr}
 	\tabcolsep=0.05cm
 	\begin{tabular}{|l|ccc|ccc|}
 \hline
 				& \multicolumn{3}{c}{BZ Cam}	  & \multicolumn{3}{c}{MV Lyr}\\
 				& \multicolumn{2}{c}{high state}  & low state & \multicolumn{2}{c}{high state}		 & low state \\
\hline
\hline
Satellite     & \multicolumn{2}{c}{\sl Swift}  & {\sl Swift}& \multicolumn{2}{c}{\sl Swift} 	  & {\sl Swift}  \\
Models     	  & Raymond-Smith 	 & apec+apec 	  & apec+apec 	& apec+plow 	 & apec+apec   	  & apec+apec 	 \\
N(H)($10^{22}$) & $0.15\pm 0.03$ & $0.29\pm 0.11$ & $<0.87$ 	  &  - 			 & $0.86 \pm 0.22$& $0.7\pm 0.6$ \\
Photon Index  &				 	 &				  &		  		  & $1.0 \pm 0.3$& 	- 			  &  - 			 \\
T$_1$ (keV)   & $9.9\pm 2.8$  	 & $11 \pm 3$     & $>10.8$ 	  & $0.7\pm0.6$  & $12 \pm 7$	  & $7 \pm 3$ 	 \\
T$_2$ (keV)   & -       		 & $0.57\pm 0.26$ & $0.5 \pm 0.4$ &  - 			 & $0.09\pm 0.03$ & $0.15 \pm 0.10$ \\ 
Flux$_{\rm{abs}}$  ($\frac{erg}{cm^{2}s}$) 		  & $3.06\times10^{-12}$ & $3.25\times10^{-12}$ & $5.34\times10^{-12}$& $5.54\times10^{-12}$ & $5.17\times10^{-12}$ & $5.29\times10^{-13}$ \\
Flux$_{\rm{unabs}}$ ($\frac{erg}{cm^{2}s}$) & $3.10\times10^{-12}$ & $3.30\times10^{-12}$ & $5.59\times10^{-12}$& $5.62\times10^{-12}$ & $5.53\times10^{-12}$ & $5.98\times10^{-13}$ \\
$\chi^2$      & 1.444 		 	 & 1.115		  & 1.004		& 1.05 			 & 0.924 		  &  - \\
\hline
\hline
\end{tabular}
\end{center}
\label{tab:bzmvmod}
\end{table*}

 V794 Aql varies between 14th and 15th magnitude in the high optical state, and in the 
 low states it can plunge to 18th-20th mag. The orbital 
period is $P_{orb}=0.1533 d.$ (Honeycutt\&Schlegel, 1985). Godon et al (2007) 
derived the following binary system parameters: $M_{WD}=0.9 M_{\odot}$, 
 high state $\dot{M} = 10^{-8.5} - 10^{-8.0} M_{\odot}/yr$, inclination $i = 60^o$, 
 and distance to the system d = 690 pc. 
 We fitted the high state spectrum of V794 Aql with two APEC components 
(see Table 2). In both components we need 
high abundance of Mg and Ni and underabundant He, S and Fe.

\begin{table*}
\footnotesize
 \begin{center}
 \caption{Fitting models and parameters for TT Ari and V794 Aql}
 \tabcolsep=0.05cm
 \begin{tabular}{|l|cccc|cc|}
 \hline
 				& 								\multicolumn{4}{c}{TT Ari} 		 		 & \multicolumn{2}{c}{V794 Aql} \\
        & \multicolumn{2}{c}{high state} & \multicolumn{2}{c}{low state}		 & high state & low state\\
\hline 
\hline
Satellite    & \multicolumn{2}{c}{\it ROSAT}   & \multicolumn{2}{c}{{\sl Swift}} 	 & {\sl Swift}& {\sl Swift}\\
Models     & Black body   & Raymond+apec   & Raymond+apec & apec & vapec+vapec & apec \\
N(H)($10^{22}$) & $<0.075$		 & $0.0265\pm0.0015$ & $0.04\pm0.03$& $0.4\pm 0.3$ & $0.05\pm0.04$& $<0.1$ \\
T$_1$ (keV)		& $0.338\pm0.017$& $0.74 \pm 0.12$   & $0.6\pm 0.3$ & $3.8 \pm 1.1$& $31\pm23$	  & $15\pm10$\\
T$_2$ (keV)		& -			 	 & $> 7.4 $		 	 & $4.6\pm 1.8$ & - & $0.97\pm0.35$& -\\
Flux$_{\rm{abs}}$ ($\frac{erg}{cm^{2}s}$)& $6.27\times10^{-13}$ 	& $1.96\times10^{-12}$&$5.26\times10^{-13}$ & $5.24\times10^{-13}$ & $5.72\times10^{-12}$ & $1.82\times10^{-12}$\\
Flux$_{\rm{unabs}}$($\frac{erg}{cm^{2}s}$)& $6.28\times10^{-13}$ 	& $1.97\times10^{-12}$&$5.28\times10^{-13}$ & $5.25\times10^{-13}$ & $5.74\times10^{-12}$ & $1.82\times10^{-12}$\\
$\chi^2$		& 1.64 		 	 & 	0.924 			 & 1.069  		& 1.219 & 1.011 				  & 0.6\\
\hline 
\hline
\end{tabular}
\end{center}
\label{tab:ttariv794mod}
\end{table*}

\section{Discussion and Conclusions}
The first conclusion that can be derived from the presented spectra is that in all the 
objects both in high and low states there is no evidence of the thermonuclear burning on 
the surface of WD. 
An important motivation for this research has been the claim by Greiner (1998) and 
Greiner et al. (2001) that some of the WD in VY Scl-type stars must be burning hydrogen 
quietly in the low state, without ever triggering thermonuclear flashes because of the 
short duration of the burning. 
We find that the predicted supersoft X-ray source (manifestation of the 
thermonuclear burning) does not appear in the low states. We also find
 another unexpected result, namely that 
the X-ray luminosity does not follow the optical/UV drop. 
Assuming that half of the gravitational energy of the matter accreted from the secondary is
released in the boundary layer through the X-ray emission and using the following 
formula from Patterson \& Raymond (1985):

\begin{equation}
\dot{M_{16}}=\frac{L_x(0.2 - 4.0 keV)}{8.7\times10^{31}M_{0.7}^{1.8}}{\rm erg~s}^{-1}
\end{equation}

In one case, BZ Cam, the X-ray
luminosity increases in the low state, and in the others, the 
decrease in X-ray flux in the low state is smaller than that in the optical and UV. 
The X-ray flux varies much less than the optical flux, by a factor of 2 to 10 during the
transition from high to low state.

The best-fit model for the 0.3-10.0 keV broad band spectra is two-component absorbed 
thermal plasma model. We do
 not find evidence of a boundary layer of the accretion disk emitting in the X-ray range. 
 Probably the emission is mainly in the far UV (e.g Godon et al., 2007).
 It appears likely that some, or 
all the X-ray flux is produced in a wind from the system: an ongoing fast wind may be the 
cause of the extended BZ Cam nebula, initially classified as a planetary nebula. 
However, TT Ari which has a $\simeq$17 minutes semi-period semi-regular variability,
which can be understood in the context of accretion, but is not explained by a wind. 

The X-ray flux may also be due to a different, and coexistent mode of accretion other than 
the disk, i.e. a 
 magnetically driven 
stream to the polar caps of an intermediate polar. The X-rays and optical flux variations anticorrelate 
only in BZ Cam, which may be due to a wind causing additional absorption and obscuring the
 luminous disk. A second explanation for the lack of correlation of UV/optical versus X-ray
 flux variations, involving magnetic accretion, appears more plausible at least for three 
 of the systems. The scenario we suggest is that the stream to the polar caps still 
 continues, at decreased rate, when the accretion disks ceases to exist.

\bigskip
\bigskip
\noindent {\bf DISCUSSION}

\bigskip
\noindent {\bf DMITRY BISIKALO:} Could you estimate the impact of bow shock to observed 
X-ray flux? 

\bigskip
\noindent {\bf POLINA ZEMKO:} The BZ Cam image in H${\alpha}$ 
(Figure 4 of Greiner et al., 2001) shows two distinct bright features in the bow shock front,
 one at $\sim$12 arcsec, the other at $\sim$40-45 arcsec
 from the central source. Nevertheless, the PSF of the 
 source in X-ray image obtained with {\sl Swift} is quite symmetric and typical of a 
 point source. So even with the modest spatial resolution of {\sl Swift} we can rule out
 that the bow shock contributes significantly to the X-ray flux.

\end{multicols}
\end{document}